\documentclass[10pt,twocolumn,letterpaper]{article}

\usepackage{wacv}
\usepackage{times}
\usepackage{epsfig}
\usepackage{graphicx}
\usepackage{amsmath}
\usepackage{amssymb}
\usepackage[ruled,vlined]{algorithm2e}
\usepackage{algorithmic}
\usepackage{multirow}

\def\wacvPaperID{873} 

\wacvfinalcopy 

\ifwacvfinal
\def\assignedStartPage{1} 
\fi

\ifwacvfinal
\usepackage[breaklinks=true,bookmarks=false]{hyperref}
\else
\usepackage[pagebackref=true,breaklinks=true,colorlinks,bookmarks=false]{hyperref}
\fi

\ifwacvfinal
\else
\fi

\begin{document}

\title{Tubular Shape Aware Data Generation for Semantic Segmentation in Medical Imaging}


\author{Ilyas Sirazitdinov$^{1,3}$,
  Heinrich Schulz$^2$,
  Axel Saalbach$^2$,
  Steffen Renisch$^2$ and
  Dmitry V. Dylov$^3$\\
 	$^1$Philips Research, Moscow, Russia\\
 	$^2$Philips GmbH Innovative Technologies, Hamburg, Germany\\
	$^3$Skolkovo Institute of Science and Technology, Moscow, Russia\\
{\tt\small ilyas.sirazitdinov@philips.com, d.dylov@skoltech.ru}\\
}

\maketitle
\begin{abstract}
Chest X-ray is one of the most widespread examinations of the human body. 
In interventional radiology, its use is frequently associated with the need to visualize various tube-like objects, such as puncture needles, guiding sheaths, wires, and catheters.
Detection and precise localization of these tube-like objects in the X-ray images is, therefore, of utmost value, 
catalyzing the development of accurate target-specific segmentation algorithms. 
Similar to the other medical imaging tasks, the manual pixel-wise annotation of the tubes is a resource-consuming process. 
In this work, we aim to alleviate the lack of the annotated images by using artificial data. 
Specifically, we present an approach for synthetic data generation of the tube-shaped objects, with a generative adversarial network being regularized with a prior-shape constraint.
Our method eliminates the need for paired image--mask data and requires only a weakly-labeled dataset (10--20 images) to reach the accuracy of the fully-supervised models.
We report the applicability of the approach for the task of segmenting tubes and catheters in the X-ray images, whereas the results should also hold for the other imaging modalities.
\end{abstract}


\section{Introduction}
Semantic segmentation is an important task in computer vision, with a plethora of applications thriving in many different domains. 
In medical imaging, it is employed to localize tissues and organs, measure their size and volume, detect pathologies, etc.~\cite{patil2013medical,kholiavchenko2020contour}.
In our work, we aim to address a clinical need to detect and localize misplaced interventional devices in X-ray images, where the segmentation could be the first step to localize the targeted object~\cite{yi2020computer}. Catheters, sheath tubes, puncture needles, and wires comprise a class of these tube-like interventional objects, the exact positioning of which \emph{in-vivo} is essential for the successful outcome of many clinical procedures~\cite{yi2020automatic,yi2020computer,frid2019endotracheal}. Endotracheal tubes, tracheostomy tubes, chest tubes, nasogastric tubes, central venous catheters, etc.\cite{yi2020computer} are all frequently engaged in clinical practice, with Chest X-ray (CXR) being used for inserting and guiding the object in a precise and a controllable manner.
Malposition of these tubes and catheters can be caused by an incorrect placement of a certain part of the tubular shape object (e.g. the tip), oftentimes leading to serious complications.

While conventional deep-learning based approaches could be employed for the localization of these devices, they usually require large volumes of manually annotated paired data~\cite{subramanian2019automated}.
Despite the widespread clinical use of catheters and tubes~\cite{yi2020automatic,yi2020computer,frid2019endotracheal}, there is a staggering shortage of such labeled paired examples available publicly, motivating us to consider a use of synthetic images alongside only a small number of labelled images to alleviate the challenge.

The main contribution of our work is the new framework for the data generation that can be trained in a weakly-supervised manner to perform segmentation of target interventional objects, relying on the prior knowledge about their tubular shape. 
More specifically, we developed a generative model that draws synthetic tubes given a pair of images: a CXR image and a fake mask for the tubular objects (Section \ref{sec:method} covers details). To achieve that, during the training, we forced the generative model to produce tubular shape objects (i.e., shape constraint) with the natural look (i.e., realistic appearance constraint) in the masked regions (Section \ref{sec:frangi}).
In the experimental Section~\ref{sec:results}, we show that this synthetic data can be useful for learning proper feature representations, yielding a better segmentation performance after fine-tuning on just a small number of the labelled examples.
\begin{figure*}[ht]
\centering
\includegraphics[width=\textwidth]{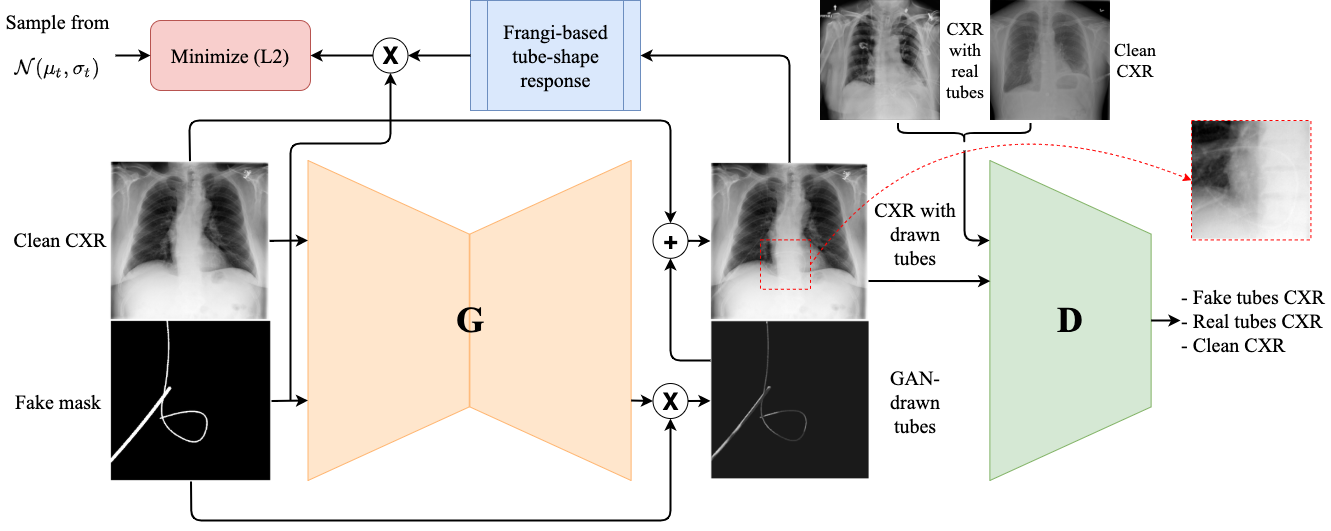}
\caption{
Proposed scheme for data generation. Three principal components are: generator (G), discriminator (D), and Frangi-based regularization block for tubular objects.
}
\label{fig:diag_synth}
\end{figure*}
\subsection{Related work}
Training models on synthetic data is a widespread methodology in machine learning. Synthetic data usually substitutes real data in cases when the real samples are inaccessible or when the annotation cost is high~\cite{chowdhury2017blood,nikolenko2019synthetic}.
Artificial data can be used in multiple ways, ranging from direct application to the model training, to the real data augmentation, to resolving the legal issues which restrict the use of the real data~\cite{nikolenko2019synthetic}. 
Medical imaging is the domain where the real annotation is especially hard to obtain due to several reasons. Manual labeling is a resource-exhaustive procedure where the usual requirements include multiple annotations of the same sample by trained professionals~\cite{wolterink2016evaluation}. Moreover, some abnormal cases might be very rare in clinical practice, making the collection of a sufficiently large (and a balanced) dataset very challenging. Finally, privacy concerns influence the collection and the release of the real clinical data publicly~\cite{choi2017generating}.

Generative adversarial networks (GAN) have been widely adopted to produce synthetic data in the medical imaging domain~\cite{beers2018high,korkinof2018high,prokopenko2019synthetic,prokopenko2019unpaired}. 
Synthetic data was also used for the semantic segmentation task, e.g. Authors of \cite{neff2018data} leveraged Wasserstein GANs with the gradient penalty to generate paired image--mask samples from the random noise. 
Authors of \cite{bailo2019red} proposed to generate synthetic masks and to transform them into realistic images so that the obtained pair can be used for the supervised segmentation. 
A narrow line of works that employs \emph{non-paired} synthetic generation of medical images, entails synthetic data from an auxiliary imaging modality in a domain adaptation setting~\cite{dong2019synthetic,prokopenko2019synthetic,prokopenko2019unpaired}; however, the segmentation performance on the synthetic data has not been evaluated in these works.
In general, one may notice that the reported methods of data generation for the image segmentation task usually require paired image--mask samples, whereas the substitute concept of using only weakly-labeled data with the proper shape constraints instead, has been dismissed in the community.

Explicit formulation of a prior shape knowledge has been widely leveraged throughout computer vision to aid the segmentation problems~\cite{heimann2009statistical,cootes1995active,ibragimov2014shape}. For example, the tubular pre-processing with the Frangi filter~\cite{frangi1998multiscale} has become a standard go-to approach to enhance the contrast of the vessels. Our work partially borrows the multi-scale Frangi filter, with the emphasis on the eigenvalue analysis of the Hessian matrix to regularize the tubular contours \cite{frangi1998multiscale,sato19973d,lorenz1997multi}. We also refer to the recent paper that proposed the \textit{clDice} loss function~\cite{shit2020clDice}, particularly efficient in training neural networks with the tubular structures.
\section{Methodology}
\begin{figure*}[h]
\centering
\includegraphics[width=\textwidth]{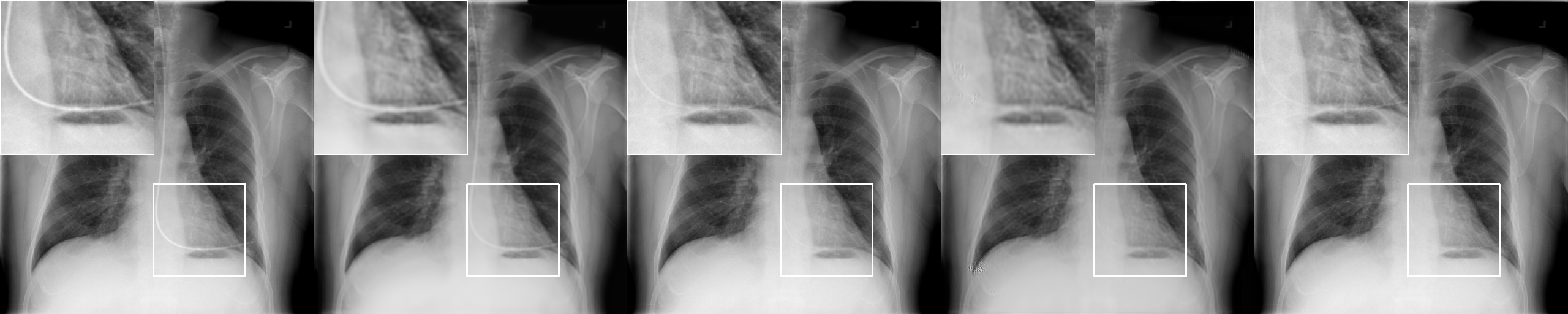}
\caption{
Different types of synthetic data compared in this work, from left to right: ``smoothed mask'', ``smoothed mask'' after the style transfer, ``hand-drawn tubes'', ``hand-drawn tubes'' after the style transfer, our method.
}
\label{fig:comp_paining}
\end{figure*}
\label{sec:method}
\subsection{Synthetic data generation}
In our work, we present a weakly-supervised synthetic data generation methodology, where we generate CXR images with the painted tube-shaped objects using only the weakly labeled information about the presence of the real tubes and catheters. Figure~\ref{fig:diag_synth} illustrates the proposed methodology. As can be seen from the figure, there are three main components: generator ($G$), discriminator ($D$), and a Frangi-based tubular regularization block. 
The generator is an encoder-decoder convolutional neural network (CNN) that takes a clean CXR and an artificial mask of the tubes as an input (a 2-channel image) and produces tube-shaped paintings, masked by the tubes. The resulting painting is summed with the initial clean CXR to obtain the CXR with the synthetic tube. 
The discriminator is a feed-forward CNN that learns to distinguish between 3 classes: CXR with the real tubes, CXR with the synthetic tubes and an additional auxiliary class of clean CXR. 
The main idea of adding the Frangi-based tube-shaped regularization is to force the generator to produce objects of tubular nature in the masked region. 
Otherwise, the generator would have drawn some random non-tubular objects in the corresponding masked regions. A detailed description of tuning the regularization block can be found in Section~\ref{sec:frangi}.
\newline
\emph{Hand-drawn tubes.} In the proposed framework, we also accompany the generated tube-containing CXR images with those drawn manually. Obviously, this kind of drawing is not an overly difficult task to perform by hand. We developed the first manual baseline data generation as a Gaussian smoothing of the fake tubes masks and further addition of the smoothed mask to the CXRs. We will refer to it as the ``smoothed mask'' method. Following Ye et. al. \cite{yi2020automatic}, we rigged the second baseline data generation process for comparing it with the GAN-drawn results. Similar to the GAN-methodology, the second baseline synthetic data generation relies on the fake masks and CXRs to generate synthetic tube intensities, conditioned on the local CXR intensity. More specifically, we randomly smooth fake masks, add borderlines and centerlines to imitate realistic tubular profile, perform random distortion and add the obtained tube image to the CXRs adapting to the local intensity. We will call it the ``hand-drawn'' method. In addition, we performed a style transfer of both baselines to decrease the fake and the real tubes domain difference. The style transfer was implemented as an unpaired image-to-image translation using Cycle GAN approach\cite{zhu2017unpaired}. Figure \ref{fig:comp_paining} shows synthetic images generated by the ``smoothed mask'' and the ``hand-drawn'' methods, their style-transfer modifications, and the images generated by our GAN.
\subsection{Frangi filter tubular regularization}
\label{sec:frangi}
The tubular constraint proposed in this work is inspired by the classical differentiable Frangi filter~\cite{frangi1998multiscale}. Frangi tubular enhancement~\cite{frangi1998multiscale} relies on the local analysis of an image $I$. Specifically, consider Taylor expansion in the neighborhood of a point $x_{0}$ which approximates the structure of $I$ up to the second order:
\begin{equation}
  I(x_0 + \delta x_0, \sigma) \approx I(x_0,\sigma) + \delta x^{T}_{0} \nabla_{0, \sigma} + \delta x^{T}_{0} \mathcal{H}_{0, \sigma} \delta x_0\,,
  \label{equ:taylor}
\end{equation}
where $\nabla_{0, \sigma}$ and $\mathcal{H}_{0, \sigma}$ are the gradient vector and the Hessian matrix at the point $x_0$ and the scale $\sigma$.
The differentiation is defined as a convolution (equation \ref{equ:conv}) with the derivatives of a D-dimensional Gaussian (equation \ref{equ:gaussian}):
\begin{equation}
    \frac{\partial}{\partial x} I(x, \sigma) = I(x) * \frac{\partial}{\partial x}G(x, \sigma)\,,
    \label{equ:conv}
\end{equation}

\begin{equation}
    G(x, \sigma) = \frac{1}{(\sigma\sqrt{2\pi})^D} e^{-\frac{\| x \|^2}{2 \sigma^2}}\,.
    \label{equ:gaussian}
\end{equation}
The main idea of the second order decomposition is to perform the eigenvalue analysis of the Hessian matrix, with the following extraction of the principal directions along which the local second order structure can be decomposed. By the eigenvalue definition, $\mathcal{H}_{0, \sigma} \hat{u}_{\sigma,k} = \lambda_{\sigma,k} \hat{u}_{\sigma,k} $, $\lambda_{\sigma,k} = \hat{u}_{\sigma,k}^T \mathcal{H}_{0, \sigma} \hat{u}_{\sigma,k}$, where $\lambda_{\sigma,k}$ is an eigenvalue of the k-$th$ normalized eigenvector $\hat{u}_{\sigma,k}^T$ of $\mathcal{H}_{0, \sigma}$ at scale $\sigma$. 
Consequently, one can observe that the eigenvalue decomposition gives 2 orthonormal directions which are invariant up to a scaling factor when mapped by the Hessian matrix. 
The obtained eigenvalues ($| \lambda_1 | \leq | \lambda_2 |$ for 2-dimensional images) can be used as an intuitive tool for modeling the geometric similarity measures, i.e., a point of a tube-shaped region should have a small value of $\lambda_1$ and a large magnitude of $\lambda_2$. As a result, the ideal tubular structure can be formulated as
\begin{equation}
    \begin{cases}
        | \lambda_1 | \approx 0\,, \\
        | \lambda_1 | \ll | \lambda_2 |\,.
    \end{cases}
    \label{equ:eigen_eq}
\end{equation}
Following~\cite{frangi1998multiscale}, we introduce two additional measures: $\mathcal{R_B}$ (equation \ref{equ:rb}) is a ``blobness'' measure (maximized for a blob-like structure) and $\mathcal{S}$ (equation \ref{equ:s}) is a measure of the ``second order structureness'' (minimized in the background, where no structure is present):
\begin{equation}
    \mathcal{R_B} = \frac{|\lambda_1|}{\sqrt{\lambda_2}}\,,
    \label{equ:rb}
\end{equation}

\begin{equation}
    \mathcal{S} = \| \mathcal{H} \|_F = \sqrt{\sum_{j \leq D} \lambda_j^2}\,.
    \label{equ:s}
\end{equation}
Both measures partake in the computation of the tube-shape response (equation \ref{equ:frangi}), where $\beta$ and $C$ are some constant thresholds:
\begin{equation}
    \mathcal{V(\sigma)} = 
    \begin{cases}
        0, & \text{if } | \lambda_1 | > 0\,, \\
        \exp\big(-\frac{\mathcal{R_B}^2}{2\beta^2}\big) \left(1 - \exp\big(-\frac{\mathcal{S}^2}{2C^2}\big)\right)\,.
    \end{cases}
    \label{equ:frangi}
\end{equation}
Following the multiscale principle, the final tube-shape response is calculated as an average of softmax among the different scales:
\begin{equation}
    \mathcal{V} = \mathbb{E} [\underset{\sigma_{\min} \leq \sigma \leq \sigma_{\max}}{\text{softmax}}\mathcal{V(\sigma)}]\,.
    \label{equ:multiscale}
\end{equation}
The calculation of the mean tube-shape response through the binary mask is summarized in Algorithm \ref{alg:tub}.
\begin{algorithm}[]
   \caption{Tube-shape response calculation}
   \label{alg:tub}
\begin{algorithmic}
   \STATE {\bfseries Input:} CXR with synthetic tubes $I$, binary mask of synthetic tubes $M$, list of different scales $\sigma$ of size $N$
   \STATE Initialize list of tube-shape responses $V$ of size $N$
   \FOR{$i = 0$ {\bfseries to} $N$}
   \STATE Calculate $\frac{\partial I}{\partial^2 x}, \frac{\partial I}{\partial x \partial y}, \frac{\partial I}{\partial^2 y}$ using second order \newline Gaussian derivatives with $\sigma_{i}$
   \STATE Compose $\mathcal{H}_{i}$ and calculate eigenvalues $\lambda_1, \lambda_2$ \newline such that $| \lambda_1 | \leq | \lambda_2 |$
   \STATE Calculate $V_i$ using equation (\ref{equ:frangi})
   \ENDFOR
   \STATE Calculate $V_{\text{multi}}$ using equation (\ref{equ:multiscale})
   \STATE Calculate $V_{\text{avg}} = \mathbb{E}_{M=1}[V_{\text{multi}} \odot M]$
   \STATE {\bfseries Return:} $V_{\text{avg}}$
\end{algorithmic}
\end{algorithm}
\begin{figure*}[h]
\centering
\includegraphics[width=\textwidth]{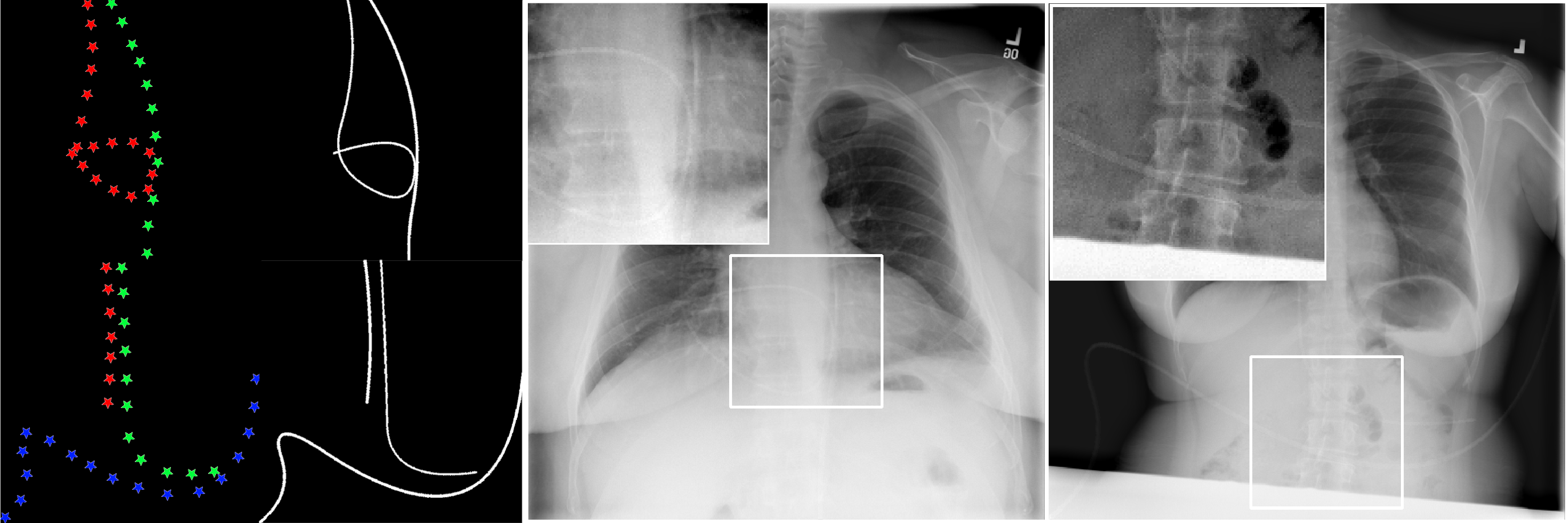}
\caption{
Fake mask generation: sampling points from an approximate location of the target objects and line interpolation ({\it left}), the resulting GAN-drawn objects ({\it middle, right}).
}
\label{fig:mask_gen}
\end{figure*}
\subsection{Loss functions} 
In our work, we use GAN strategy to train a model for synthetic data generation. The conventional definition of GAN learning is a two-player game, where a discriminator ($D$) and a generator ($G$) compete with each other. In our problem, this definition can be formulated as a min-max game (Eq. \ref{eq:GAN}):
\begin{equation}
    \begin{aligned}
    \min\limits_{G}\max\limits_{D}V(G,D) = \mathbb{E}_{{x}\sim{p_{\text{data}}}}[\log{D({x})}]+\\
    \mathbb{E}_{{z}\sim{\text{noise}}}[\log{(1-D(G({z})))}]
    \label{eq:GAN}
    \end{aligned}
\end{equation}
Recall that we aim to learn a mapping from the clean CXR ($X_{cl}$), conditioned on the tube masks ($m$), to the CXR images with the real tubes ($X_{r}$). After extensive experimentation with the conventional GAN loss functions, we stopped at the least squares GAN (LSGAN) \cite{mao2017least} with additional class which can be written as:
\begin{equation}
    \begin{aligned}
    \mathcal{L}_{\text{LSGAN}}(D) = \frac{1}{3}\Big[&\mathbb{E}_{x\sim{p(X_{r})}}(D(x)-a)^2+\\
    &\mathbb{E}_{x\sim{p(X_{cl})}}\big(D(x) - b\big)^2+\\
    &\mathbb{E}_{x\sim{p(X_{cl})}}\big(D(x+ m \odot G(x,m)) - c\big)^2\Big],\\
    \mathcal{L}_{\text{LSGAN}}(G) = \frac{1}{3}\Big[&\mathbb{E}_{x\sim{p(X_{cl})}}\big(D(x+ m \odot G(x,m))-a\big)^2\Big],
    \end{aligned}
    \label{eq:LSGAN}
\end{equation}
where $a$, $b$, $c$ are one-hot vectors of three possible classes: $a$ corresponds to the class of CXRs with the real tubes, $b$ is for class of the clean CXRs, $c$ is for the class of the synthetic tubes.
The tube-shape constraint can also be defined as an additional loss component $\mathcal{L}_{tub}$. If we denote the differentiable Algorithm \ref{alg:tub} as $F(x,m, \sigma_N)$, then $\mathcal{L}_{tub}$ can be formulated as:
\begin{equation}
    \mathcal{L}_{tub} = \big(F(x,m, \sigma_N) - t\big)^2,
    \label{eq:ltub}
\end{equation}
where $\sigma_N$ is a list of sigmas for the Frangi-filter, $t$ is a sample from the mean tube-shape response distribution with parameters $\mu_t, \sigma_t$ (estimated on few images of CXR images with the real tubes), i.e., $t \sim \mathcal{N}(\mu_t, \sigma_t)$.
Given the definitions of the adversarial loss and the tube-shape loss, we can formulate our final objective function as:
\begin{equation}
    \begin{aligned}
    &D^{*} = arg\min\limits_{D}\Lambda_{1}\mathcal{L}_{\text{LSGAN}}(D),\\
    &G^{*} = arg\min\limits_{G}(\Lambda_{1}\mathcal{L}_{\text{LSGAN}}(G)+\Lambda_{2}\mathcal{L}_{tub}),
    \end{aligned}
    \label{eq:total}
\end{equation}
where $\Lambda_{1}, \Lambda_{2}$ are the hyperparameters that weight the relative contribution of each component in equations (\ref{eq:total}).
\subsection{Segmentation task formulation given synthetic data}
In our work, we synthesize paired samples $X_{\text{synth}}$ and $M_{\text{synth}}$, where $X_{\text{synth}}$ is a set of CXR images with the generated synthetic tubes, and $M_{\text{synth}}$ is a set of the binary masks of these tubes. 
The pairs are then used to train a segmentation model in the supervised manner. We train the segmentation model using the generated \emph{synthetic} pairs and validate it on the CXR images with \emph{the real} tubes, wires and catheters. We then fine-tune the trained models on a small number of labelled real images both to show the applicability of the synthetic data and to inhibit the domain shift~\cite{choi2017generating,prokopenko2019synthetic}.

\section{Results}
\label{sec:results}
\begin{table*}[h]
\caption{Comparison of models trained on Manual (Hand-drawn) and Generated (GAN-drawn) synthetic data. Base.1, 2 corresponds to baseline 1,2. S.T. indicates whether the dataset was adapted using a style transfer. Bold font shows statistically significant best result. BCE corresponds to basic binary cross-entropy loss.}
\begin{center}
	\begin{tabular}{c|c|ccccc}
		\textbf{Loss}                       & \textbf{Type} & \textbf{SoftDice}     & \textbf{Dice}         & \textbf{Precision}    & \textbf{Recall}       & \textbf{Hausdorf}      \\
		\hline \hline
		\multirow{2}{*}{\textbf{BCE,1}}     & Base.1        & 6.8$\pm$2.4           & 6.0$\pm$2.1           & 51.4$\pm$5.1          & 3.4$\pm$1.4           & 8.28$\pm$0.06          \\
		                                    & Base.1+S.T.        & 14.3$\pm$1.7 &	13.2$\pm$1.9 &	68.5$\pm$1.7 &	7.9$\pm$1.4	& 8.15$\pm$0.07          \\
		                                    & Base.2.        & 40.0$\pm$0.9          & 42.2$\pm$1.1          & 81.5$\pm$1.2 & 31.2$\pm$1.0          & 7.14$\pm$0.21          \\
		                                    & Base.2+S.T.        & 40.4$\pm$0.4 &	44.1$\pm$0.5 & 80.5$\pm$0.6 & 32.9$\pm$0.4 & 7.18$\pm$0.05  \\
		                                    & Ours GAN           & \textbf{41.7$\pm$0.9} & \textbf{46.0$\pm$0.8} & 78.0$\pm$1.2          & \textbf{35.6$\pm$0.7} & \textbf{7.11$\pm$0.04} \\
		\hline
		\multirow{2}{*}{\textbf{BCE,8}}     & Base.1        & 11.0$\pm$0.6          & 10.7$\pm$0.6          & 49.6$\pm$1.6          & 6.6$\pm$0.5           & 8.20$\pm$0.05          \\
		                                    & Base.1+S.T.        & 22.9$\pm$2.5 &	24.3$\pm$3.2 &	60.2$\pm$1.5 &	17.3$\pm$2.7 &	8.34$\pm$0.054 \\
		                                    & Base.2        & 44.9$\pm$0.6 & 51.7$\pm$1.1          & \textbf{62.3$\pm$1.1} & 51.7$\pm$3.0          & \textbf{7.23$\pm$0.29} \\
		                                    & Base.2+S.T.        & 44.4$\pm$1.1 & 53.8$\pm$0.5 & 55.5$\pm$2.0 &	59.6$\pm$0.9 & 7.54$\pm$0.14 \\
		                                    & Ours GAN           & 41.7$\pm$1.9          & 51.8$\pm$2.0          & 51.4$\pm$2.1          & \textbf{60.8$\pm$1.1} & 7.9$\pm$1.0            \\
		\hline
		\multirow{2}{*}{\textbf{BCE, Dice}} & Base.1        & 6.9$\pm$1.5           & 6.7$\pm$1.5           & 47.6$\pm$4.0          & 3.9$\pm$1.0           & 8.28$\pm$0.6           \\
		                                    & Base.1+S.T.        & 15.7$\pm$2.7 & 15.5$\pm$2.8 & 67.3$\pm$2.1 &	9.6$\pm$2.0 & 8.09$\pm$0.7 \\
		                                    & Base.2        & 46.0$\pm$1.6          & 47.1$\pm$1.5          & 76.2$\pm$2.5 & 38.1$\pm$1.0          & 7.02$\pm$0.13          \\
		                                    & Base.2+S.T.        & 50.4$\pm$1.1 &	51.8$\pm$1.4 &	74.8$\pm$2.5 &	43.5$\pm$2.6 &	7.01$\pm$0.03 \\
		                                    & Ours GAN           & 51.2$\pm$0.9 & \textbf{53.8$\pm$0.7} & 67.4$\pm$1.1          & \textbf{49.6$\pm$1.1} & 7.07$\pm$0.23          \\
		\hline
		\multirow{2}{*}{\textbf{clDice}}    & Base.1        & 6.2$\pm$1.5           & 6.2$\pm$1.5           & 43.1$\pm$4.8          & 3.7$\pm$1.0           & 8.32$\pm$0.07.         \\
		                                    & Base.1+S.T.        & 18.6$\pm$2.0 &	18.6$\pm$2.0 &	63.6$\pm$2.2 & 12.1$\pm$1.5 & 8.07$\pm$0.05 \\
		                                    & Base.2        & 44.7$\pm$3.6          & 44.7$\pm$3.6          & 72.6$\pm$3.6 & 35.7$\pm$3.7          & 7.06$\pm$0.13          \\
		                                    & Base.2+S.T.        & 52.1$\pm$0.8 &	52.2$\pm$0.9 &	72.0$\pm$0.5 &	44.8$\pm$1.1 &	7.04$\pm$0.6
 \\
		                                    & Ours GAN           & 52.8$\pm$0.6 & 52.8$\pm$0.7 & 67.6$\pm$1.0          & \textbf{47.7$\pm$0.9} & 7.21$\pm$0.46          
	\end{tabular}
\end{center}
\label{tab:man_gan}
\end{table*}
\begin{table*}[h]
\caption{Tuning models on small amount of labelled data. ``\emph{De novo}'' corresponds to initialization from scratch (with a model pre-trained on ImageNet),``Base.'' corresponds to baselines models, ``S.T.'' corresponds to additional style transfer processing, ``GAN'' corresponds to the best validation parameters on the GAN-drawn synthetic data. Bold font emphasizes statistically significant best result.}
\begin{center}
  \begin{tabular}{c|c|ccccc}
    \textbf{Loss}                   & \textbf{Method} & \textbf{SoftDice}     & \textbf{Dice}         & \textbf{Precision}    & \textbf{Recall}       & \textbf{Hausdorf}      \\
    \hline \hline
    & & \multicolumn{5}{c}{\textbf{10 images}} \\
    \multirow{3}{*}{\textbf{BCE,1}}       & \emph{De novo}         & 31.0$\pm$3.3          & 48.4$\pm$1.7          & 61.5$\pm$2.8          & 44.0$\pm$1.2          & 7.42$\pm$0.14          \\
                                    & Base.1          & 53.$\pm$0.4          & 55.6$\pm$0.2          & 68.9$\pm$0.5          & 50.4$\pm$0.0          & 6.90$\pm$0.02          \\
                                    & Base.1+S.T.          & 54.1$\pm$0.4 &	56.2$\pm$0.4 & 73.7$\pm$0.6 & 49.2$\pm$0.6 & 6.81$\pm$0.04 \\
                                    & Base.2          & 59.2$\pm$0.2          & 62.4$\pm$0.1          & 74.0$\pm$0.4          & 58.1$\pm$0.2          & 6.58$\pm$0.01          \\
                                    & Base.2 + S.T. & 59.0$\pm$0.5 & 61.0$\pm$0.5 &	71.7$\pm$0.6 &	57.6$\pm$0.2 &	6.55$\pm$0.02 \\
                                    & GAN             & \textbf{61.0$\pm$0.0} & \textbf{63.9$\pm$0.1} & \textbf{75.2$\pm$0.5} & \textbf{59.5$\pm$0.4} & \textbf{6.48$\pm$0.02}          \\
    \hline
    \multirow{3}{*}{\textbf{clDice}}      & \emph{De novo}         & 49.1$\pm$0.6          & 51.9$\pm$0.4          & 63.4$\pm$1.7          & 48.2$\pm$1.5          & 7.32$\pm$0.07          \\
                                    & Base.1.          & 55.7$\pm$1.1          & 55.9$\pm$1.1          & 70.5$\pm$0.7 & 49.9$\pm$2.0          & 6.93$\pm$0.06          \\
                                    & Base.1+S.T.         & 57.1$\pm$0.1 & 57.3$\pm$0.1 & 72.7$\pm$0.2 & 51.2$\pm$0.1 & 6.81$\pm$0.05 \\
                                    & Base.2          & 61.8$\pm$0.1          & 62.4$\pm$0.1          & 74.5$\pm$0.9          & 57.7$\pm$0.5          & 6.6$\pm$0.0            \\
                                    & Base.2+S.T.          & 61.4$\pm$0.6 &	61.6$\pm$0.6 &	73.7$\pm$1.5 &	56.9$\pm$0.7 &	6.59$\pm$0.03 \\
                                    & GAN             & \textbf{64.1$\pm$0.2} & \textbf{64.3$\pm$0.2} & 74.4$\pm$0.1          & \textbf{60.9$\pm$0.4} & \textbf{6.49$\pm$0.02} \\
    \hline \hline
    & & \multicolumn{5}{c}{\textbf{20 images}} \\
    \multirow{3}{*}{\textbf{BCE,1}}       & \emph{De novo}         & 31.4$\pm$5.6          & 51.8$\pm$1.7          & 68.2$\pm$2.7          & 45.2$\pm$2.2          & 7.21$\pm$0.12          \\
                                    & Base.1          & 54.0$\pm$0.4          & 57.0$\pm$0.4          & 74.7$\pm$0.9          & 50.1$\pm$0.4          & 6.76$\pm$0.26          \\
                                    & Base.1+S.T.          & 56.2$\pm$0.4 &	58.9$\pm$0.5 &	76.7$\pm$1.1 &	51.6$\pm$1.1 &	6.69$\pm$0.03 \\
                                    & Base.2          & 60.1$\pm$0.0          & 64.0$\pm$0.0          & 77.7$\pm$0.2          & 57.9$\pm$0.1          & 6.46$\pm$0.0           \\
                                    & Base.2 + S.T. & 61.3$\pm$0.2 & 63.7$\pm$0.4 &	\textbf{79.0$\pm$0.5} &	56.9$\pm$0.2 &	6.49$\pm$0.07 \\
                                    & GAN             & 60.9$\pm$0.5          & 64.6$\pm$0.6          & 78.3$\pm$0.1 & \textbf{58.6$\pm$0.7}          & 6.45$\pm$0.02          \\
    \hline
    \multirow{3}{*}{\textbf{clDice}}      & \emph{De novo}         & 53.8$\pm$0.7          & 56.2$\pm$0.9          & 71.3$\pm$0.4          & 50.1$\pm$1.1          & 6.94$\pm$0.03          \\
                                    & Base.1          & 57.5$\pm$0.5          & 57.6$\pm$0.5          & 72.3$\pm$1.0 & 52.2$\pm$1.0          & 6.77$\pm$0.06 \\
                                    & Base.1+S.T.         & 58.4$\pm$0.2 & 58.6$\pm$0.2 & 75.5$\pm$0.9 & 51.8$\pm$0.2 & 6.73$\pm$0.04 \\
                                    & Base.2          & 63.2$\pm$0.2          & 63.9$\pm$0.2          & 76.3$\pm$0.2          & 58.8$\pm$0.4          & 6.51$\pm$0.03 \\
                                    & Base.2+S.T.          & 62.5$\pm$0.6 &	62.7$\pm$0.6 &	\textbf{77.5$\pm$0.2} &	56.6$\pm$1.0 &	6.53$\pm$0.07 \\
                                    & GAN             & \textbf{65.3$\pm$0.1} & \textbf{65.6$\pm$0.2} & 75.6$\pm$0.5          & \textbf{61.9$\pm$0.4}          & \textbf{6.46$\pm$0.0}           
  \end{tabular}
\end{center}
\label{tab:tune}
\end{table*}
\subsection{Dataset}
We used one of the largest publicly available chest X-ray datasets in our experiments (ChestX-ray14~\cite{wang2017chestx}). We randomly selected 1850 images and manually labeled them as either ``clean'' CXRs (1120) $X_{\text{clean}}$ or as the CXRs with at least one tube, wire, or catheter present (730), i.e., each CXR received a binary label to flag the presence of a tube-shape object in the image. In addition, 200 of such tube-containing images were further manually annotated with the binary masks to indicate pixels which belong to tubes, wires, or catheters in CVAT software~\cite{cvat}.
\begin{table*}[h]
\caption{Comparison of modifications for synthetic data generation, ``Intensity'' corresponds to the mean color response regularization, ``Frangi'' to the pipeline of Fig. \ref{fig:diag_synth}, ``Frangi + Cycle'' to  Fig. \ref{fig:diag_synth} augmented with an additional generator for the cycle-consistency.} 
\begin{center}
  \begin{tabular}{c|c|ccccc}
    Loss & Method & \textbf{SoftDice}     & \textbf{Dice}         & \textbf{Precision}    & \textbf{Recall}       & \textbf{Hausdorf} \\
    \hline \hline
    \multirow{3}{*}{\textbf{BCE, 1}}    & Intensity        & 27.0$\pm$4.3          & 27.1$\pm$4.7          & 74.7$\pm$4.1          & 18.2$\pm$3.6          & 7.72$\pm$0.24     \\
                                 & Frangi + Cycle & 34.6$\pm$3.6          & 36.5$\pm$4.5          & 75.0$\pm$2.0          & 26.7$\pm$4.2          & 7.43$\pm$0.32     \\
                                 & Frangi         & \textbf{41.7$\pm$1.9} & \textbf{46.0$\pm$2.6} & \textbf{78.0$\pm$2.2} & \textbf{35.6$\pm$2.7} & 7.11$\pm$0.13     \\
    \hline
    \multirow{3}{*}{\textbf{BCE, 8}}    & Intensity        & 38.7$\pm$3.3          & 42.4$\pm$3.8          & \textbf{65.6$\pm$3.2} & 35.8$\pm$3.6          & 8.08$\pm$0.54     \\
                                 & Frangi + Cycle & \textbf{42.3$\pm$1.2} & 48.8$\pm$1.9          & 59.5$\pm$1.2          & 47.2$\pm$3.5          & 7.52$\pm$0.2      \\
                                 & Frangi         & 41.7$\pm$1.9          & \textbf{51.8$\pm$2.0} & 51.4$\pm$2.1          & \textbf{60.8$\pm$1.5} & 7.9$\pm$1.0       \\
    \hline
    \multirow{3}{*}{\textbf{BCE, Dice}} & Intensity        & 34.1$\pm$3.0          & 34.3$\pm$3.2          & \textbf{76.3$\pm$1.6} & 24.6$\pm$3.1          & 7.69$\pm$0.28     \\
                                 & Frangi + Cycle & 41.0$\pm$1.9          & 42.0$\pm$2.0          & 69.3$\pm$2.1          & 33.7$\pm$2.1          & 7.39$\pm$0.14     \\
                                 & Frangi         & \textbf{51.2$\pm$2.4} & \textbf{53.8$\pm$2.5} & 67.4$\pm$1.1          & \textbf{49.6$\pm$3.6} & 7.07$\pm$0.23     \\
    \hline
    \multirow{3}{*}{\textbf{clDice}}   & Intensity        & 31.0$\pm$4.2          & 31.0$\pm$4.2          & 70.8$\pm$2.8          & 21.6$\pm$3.5          & 7.87$\pm$0.18     \\
                                 & Frangi + Cycle & 43.2$\pm$3.6          & 43.2$\pm$3.6          & 70.2$\pm$2.0          & 34.3$\pm$3.8          & 7.42$\pm$0.14     \\
                                 & Frangi         & \textbf{52.8$\pm$0.6} & \textbf{52.8$\pm$0.7} & 67.6$\pm$1.0          & \textbf{47.7$\pm$0.9} & 7.21$\pm$0.46
  \end{tabular}
\end{center}
\label{tab:ablation}
\end{table*}
\begin{figure*}[h]
\centering
\includegraphics[width=\textwidth]{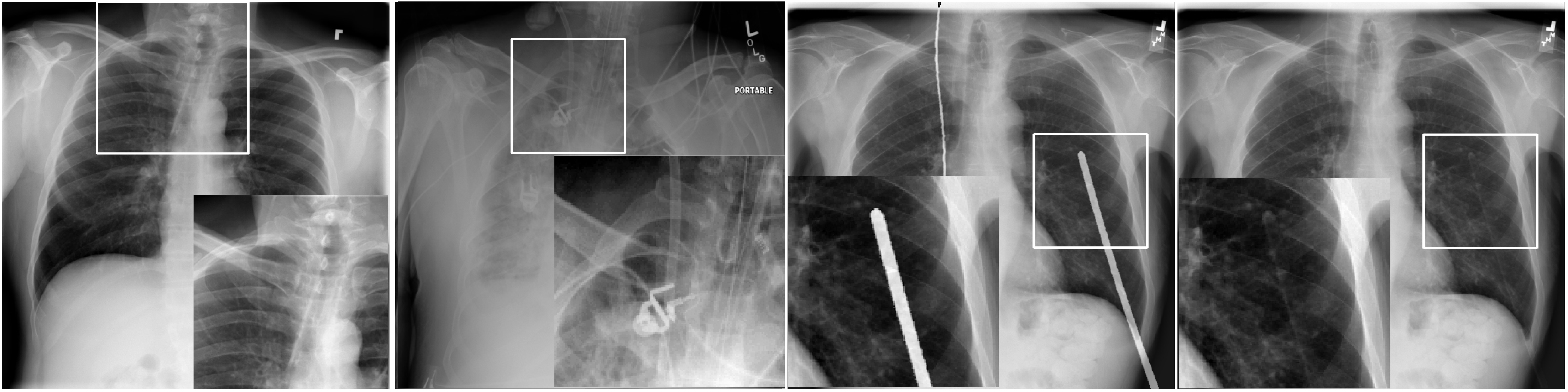}
\caption{
From left to right: an example of chest X-ray images with synthetic tubes, chest X-ray images with the real tube-shape objects, artifacts for the intensity regularization attempt  and the identical drawing by the Frangi-based model.
}
\label{fig:fake_vs_real}
\end{figure*}
Traditional supervised segmentation on synthetic data requires paired image--mask samples $X_{\text{synth}}$ and $M_{\text{synth}}$. In our work, we produce $M_{\text{synth}}$ in the following way: for each ``clean'' CXR from $X_{\text{clean}}$, we purposely generated a fake mask image with \emph{some} tubular objects in it.
This step corresponds to defining \emph{the location priors} of the particular types of tubes and catheters that we needed to generate for the clinical need, namely, the central venous catheter (CVC), the chest tube, and the endotracheal tube~\cite{yi2020computer}.
Given the type of the object, we generated the fake masks relying on the information about the possible object location, the curvature of the tube, and the size (see Fig. \ref{fig:mask_gen} for the example of a fake mask), namely we sampled points from the approximate target object location, connected them using the spline interpolation, and varied the width by tuning the operation of morphological dilation.
\subsection{Synthetic data generation}
A pair of networks (generator $G$ and discriminator $D$) with the tubular shape constraint was employed to generate synthetic data to consequently perform the segmentation. 
The generator was a U-Net-based CNN~\cite{ronneberger2015u} with the residual blocks in the encoder. 
The discriminator was a CNN of 4 stacked convolutional layers and a single fully-connected layer. 
Similarly to DCGAN \cite{radford2015unsupervised} and CycleGAN \cite{zhu2017unpaired}, we used instance normalization, LeakyReLu as the intermediate activation function, and the strided convolution for downsampling.
In addition, we applied spectral normalization~\cite{miyato2018spectral} to stabilize training process and to augment discriminator learning with the extra ``clean'' CXR class. 
Frangi-filter parameters were $\sigma_N = [2, 4, 6]$, $\beta=0.5$, $C=0.5$. The final objective function (Eq. \ref{eq:total}) with $\Lambda_1=1, \Lambda_2=10$ was minimized by Adam optimizer ($lr_{D}=6 \times 10^{-4}$, $lr_{G}=2 \times 10^{-4}$, $\beta_1 = 0.5, \beta_2 = 0.999$) for 40 epochs.
\subsection{Segmentation from synthetic data}
The segmentation pipeline included supervised training of the U-Net model with an ResNet-34 encoder pretrained on ImageNet \cite{he2016deep}. We trained the network with the input resolution of 512$\times$512 using AdamW optimizer for 60 epochs. To perform a fair evaluation, the model was trained on a series of conventional segmentation losses, namely weighted binary cross-entropy (BCE), a sum of the BCE with the Dice loss, and Dice plus clDice loss \cite{shit2020clDice}, designed to improve the detection quality of the tube-shape objects. 
The segmentation models were evaluated on a 5-fold cross-validation using the soft Dice and Dice scores, Hausdorf distance, pixel-wise precision, and recall metrics.
Table \ref{tab:man_gan} compares the segmentation given the synthetic data from two synthetic datasets: a ``manual'' (hand-drawn) and a GAN-drawn as the result of inference of the trained model (Figure \ref{fig:diag_synth}). Both models employed identical pairs of clean CXRs and fake masks. 
Remarkably, the highest Dice score ($\sim$53-54\%) was obtained using the GAN-generated synthetic samples. In the table, the relatively low Dice score could be explained by a low recall, e.g. when the highest recall score is about 60\% even for BCE with pos. weight 8 (it is strongly biased to recall).
Results for the fine-tuned models, pre-trained on synthetic data and updated on several real labels, are presented in Table \ref{tab:tune}.
\begin{figure*}[ht]
\centering
\includegraphics[width=\textwidth]{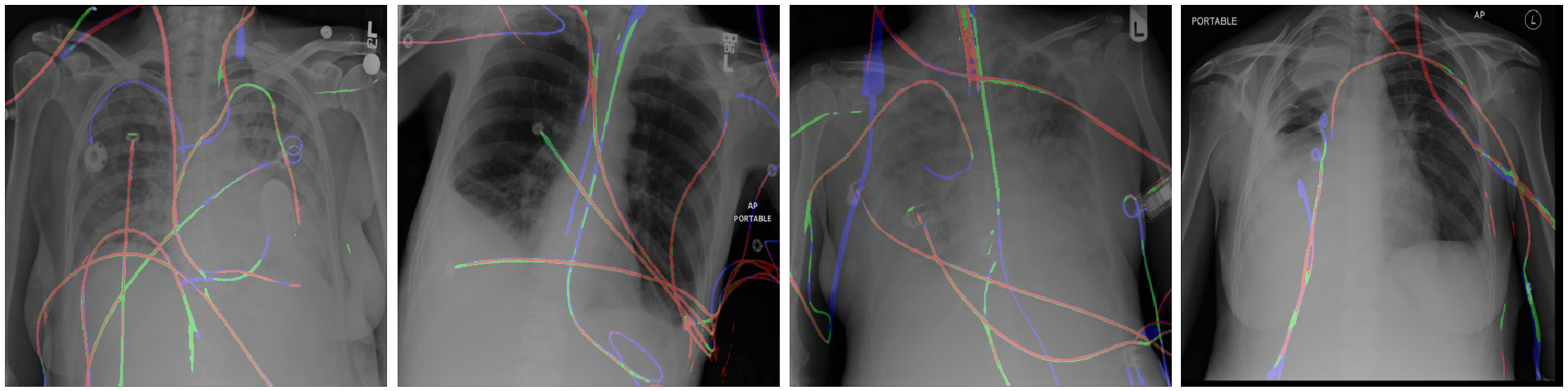}
\caption{
Segmentation results: predictions after training only on synthetic data ({\it red}), refinement after fine-tuning on small amount of ground truth data ({\it green}), missed ground truth tubes - false negative error ({\it blue}). The target objects on the chest X-ray images are tubes, catheters, wires, ECG cords, etc.
}
\label{fig:comp_segm}
\end{figure*}
\label{sec:altern}
In our experiments, we picked the model with the highest Dice on the validation set and fine-tuned it using only 5, 10 or 20 images containing the real tubes, wires, or catheters. 
In addition, we trained the segmentation model from scratch to understand the impact of the synthetic data. We report the corresponding Dice score improvement from $\sim$53-54\% to $\sim$65-66\%, and a $\sim$15-20\% gain in recall. 
\subsection{Alternative hypothesis for synthetic data generation}
This section aims to factor out contributions of the loss function terms and contains results for alternative natural ideas for generating tubular objects. 
Alternatively to the Frangi-based recipe, we could substitute Algorithm \ref{alg:tub} with a simple constraint that will calculate a mean color response in the masked region (i.e. intensity regularization), namely $F(x,m) = \mathbb{E}_{m=1}(x \odot m)$, where $x$ is a CXR image with synthetic tubes and $m$ is the corresponding mask. Then, Eq. \ref{eq:ltub} remains the same but $t$ becomes an expected color of the tube-shape objects. Similar to Frangi-based regularization, intensity regularization forces generator to produce something with a specified mean color in the masked regions. 
The second alternative idea assumes plugging in the second generator $G_{2}$ in the Cycle-GAN manner \cite{zhu2017unpaired}. In such setting, the generator $G_{2}$ takes an CXR with the fake tubes and returns the reconstructed initial CXR along with the predicted mask. In fact, the objective \ref{eq:total} is augmented with an one-sided cycle-consistency loss term: 
\begin{equation}
\begin{aligned}
    \mathcal{L}_{\text{cyc}} = \mathbb{E}_{x \sim p(X_{cl})}\Big[\|G_{2_{\text{rec}}}(x + m \odot G(x)) - x\|_{1} + \\ \mathcal{L}_{s}\big(G_{2_{\text{segm}}}(x + m \odot G(x)), m\big)\Big],
\end{aligned}
\label{eq:cycle}
\end{equation}
where $G_{2_{\text{rec}}}$ and $G_{2_{\text{segm}}}$ are the reconstruction and the segmentation mappings of $G_2$, and $\mathcal{L}_s$ is an arbitrary segmentation loss. We assumed that the cycle-based segmentation would help to produce realistic appearance of tubular shape objects in the learned regions. The results of both alternatives are presented in Table \ref{tab:ablation}. Albeit intuitive, both ideas proved inferior to our baseline model presented in Figure \ref{fig:diag_synth}, suggesting the strict Frangi-based regularization as the go-to solution.

\section{Discussion}
We presented the first GAN pipeline for the task of synthetic data generation with the tubular priors, and demonstrated its efficacy for segmenting catheters and wires in medical imaging. Our approach proved useful in mitigating the shortage of annotated data of these tubular objects, with just a dozen of labeled images proving sufficient to train a highly efficient model.
Clinical relevance can be seen in Figure \ref{fig:mask_gen} where the synthetic drawing looks indiscernible from the real tube-containing CXRs. A pair of a clean CXR and an artificial mask, fed into the model, yield a smooth, realistic, and a naturally looking radiograph.
Figure \ref{fig:comp_paining} shows the drawing quality of the ``smoothed mask'' and ``hand-drawn tubes'' methods and of our method. 
As can be noticed, our method is far more similar in appearance with the hand-drawn method than with the smoothed mask method which has higher contrast and unnatural appearance, i.e. compare with the second image in Figure \ref{fig:fake_vs_real}. 

Remarkably, we achieved this kind of quality automatically by learning from the data, while the hand-drawn method had to use predefined manual rules for drawing.
Figure \ref{fig:fake_vs_real} compares two randomly chosen images: one with an artificial tube and the other one with the real tube-shape objects. As can be noticed, the general appearance of the objects is alike; however, attention to the detail reveals minor traces of artificial nature, such as the pixelated tube borders and affected centerline of the catheters.
Despite these visual effects, we carefully examined the pairs of synthetic data samples with regard to their contribution to the supervised segmentation task. Our study (see Table \ref{tab:man_gan}) supports that the GAN-drawn data was comparable to the hand-drawn in terms of the segmentation performance. Although the generated CXRs were not able to cover the distribution of the images with real tubular objects (Dice score did not exceed 53\%), simple domain adaptation via fine-tuning the hyperparameters of our model gave a serious boost to all metrics (see Table \ref{tab:tune}). Specifically, the Dice score improved anywhere from 7 to 13\%, depending on the number of images used for fine-tuning.
Moreover, we observe that the synthetic images helped learn proper feature representations, because these fine-tuned models outperformed the ones trained from scratch by a large margin (e.g., the Dice score was better by $\sim$ 9-20\%).
Remarkably, when we compare the fine-tuned models to the ones trained in a fully supervised manner on the real data (Table \ref{tab:gt}), we observe that the segmentation quality of the fine-tuned models lags by only $\sim5\%$, whereas we need \emph{seven times less} of the real labels to perform on this level.
\begin{table*}[h]
  \caption{Comparison of segmentation models trained on the real data. Evaluation was performed on the out-of-fold predictions.}
  \begin{center}
    \begin{tabular}{c|c|ccccc}
      \textbf{Loss}   & \textbf{SoftDice} & \textbf{Dice} & \textbf{Precision} & \textbf{Recall} & \textbf{Hausdorf} \\
      \hline \hline
        \textbf{BCE,1}  & 59.8$\pm$2.2 & 68.3$\pm$2.7 & 79.4$\pm$1.7 & 62.9$\pm$3.1 & 6.29$\pm$0.65 \\
        \hline
        \textbf{BCE,8}  & 55.8$\pm$3.2 & 63.5$\pm$2.7 & 56.2$\pm$4.2 & 78.4$\pm$2.5 & 7.45$\pm$1.19 \\
        \hline
        \textbf{clDice} & 70.0$\pm$2.0 & 70.2$\pm$2.0 & 74.3$\pm$3.1 & 69.7$\pm$2.4 & 6.38$\pm$5.51 
    \end{tabular}
  \end{center}
  \label{tab:gt}
\end{table*}

We also found our Frangi-based tubular object regularization to be an efficient constraint to preserve the shape of the target objects in the synthetic data, with the alternative constraints (Section \ref{sec:altern}) not yielding any improvement. For example, Figure \ref{fig:fake_vs_real} shows a typical artifact of the mean color response regularization, where the masked regions have unnatural color. In contrast, the Frangi-based regularization produced smoothed drawing for the identical input. The results of the additional cycle loss were slightly better and outperformed the network without cycle consistency in some settings (Table \ref{tab:ablation}); yet, the Frangi-based model was better in the larger number of settings. The segmentation results are presented in Figure \ref{fig:comp_segm}. 
As can be noticed, the major part of the segmented regions ({\it red}) comes from the model pretrained on synthetic data, with the fine-tuning ({\it green}) helping to fill the regions missed initially. 
One can observe the missed false-negative tubes ({\it blue}) mostly in the area of lungs where the opacity is the highest (white color). 
This fact can be explained by the natural domain shift between the synthetic and the real CXRs. 
Our synthetic images happened to be drawn mostly on normal CXRs with the clear lungs, whereas the real CXRs do contain more irregularities (cases with pathologies and lungs with ``abnormal'' morphologies and textures). The relevant domain adaptation work to further reduce the gap between the synthetic and the real images will be shown elsewhere.

\section{Conclusion}
In this work, we presented the first method for weakly-supervised synthetic data generation of tubular objects. We showed that the synthetic data was useful for learning good feature representations, helping to achieve better segmentation scores when the models are fine-tuned on a small number of labeled images. The applicability of our synthetic data generation method was evaluated on the task of segmentation of the interventional tubes (sheath tubes, catheters, and wires). 
However, our method is not limited to any specific modality or application and can be used to address similar tasks beyond radiology, e.g., for segmentation of the vasculature in histopathology or in ophthalmology, for detection of the road marking in the autonomous vehicle video streams, or for segmenting streets/rivers in the satellite data.
Moreover, the tube-shape regularization can be extended to 3D~\cite{frangi1998multiscale}, which may help to generate synthetic volumes with the GAN-drawn tubular object (e.g., for vessel segmentation in computed tomography and beyond).
{\small
\bibliographystyle{ieee_fullname}
\bibliography{egbib}

\begin{thebibliography}{10}\itemsep=-1pt

\bibitem{bailo2019red}
Oleksandr Bailo, DongShik Ham, and Young Min~Shin.
\newblock Red blood cell image generation for data augmentation using
  conditional generative adversarial networks.
\newblock In {\em Proceedings of the IEEE Conference on Computer Vision and
  Pattern Recognition Workshops}, pages 0--0, 2019.

\bibitem{beers2018high}
Andrew Beers, James Brown, Ken Chang, J~Peter Campbell, Susan Ostmo, Michael~F
  Chiang, and Jayashree Kalpathy-Cramer.
\newblock High-resolution medical image synthesis using progressively grown
  generative adversarial networks.
\newblock {\em arXiv preprint arXiv:1805.03144}, 2018.

\bibitem{choi2017generating}
Edward Choi, Siddharth Biswal, Bradley Malin, Jon Duke, Walter~F Stewart, and
  Jimeng Sun.
\newblock Generating multi-label discrete patient records using generative
  adversarial networks.
\newblock {\em arXiv preprint arXiv:1703.06490}, 2017.

\bibitem{chowdhury2017blood}
Aritra Chowdhury, Dmitry~V Dylov, Qing Li, Michael MacDonald, Dan~E Meyer,
  Michael Marino, and Alberto Santamaria-Pang.
\newblock Blood vessel characterization using virtual 3d models and
  convolutional neural networks in fluorescence microscopy.
\newblock In {\em ISBI 2017}, pages 629--632. IEEE, 2017.

\bibitem{cootes1995active}
Timothy~F Cootes, Christopher~J Taylor, David~H Cooper, and Jim Graham.
\newblock Active shape models-their training and application.
\newblock {\em Computer vision and image understanding}, 61(1):38--59, 1995.

\bibitem{dong2019synthetic}
Xue Dong, Yang Lei, Sibo Tian, Tonghe Wang, Pretesh Patel, Walter~J Curran,
  Ashesh~B Jani, Tian Liu, and Xiaofeng Yang.
\newblock Synthetic mri-aided multi-organ segmentation on male pelvic ct using
  cycle consistent deep attention network.
\newblock {\em Radiotherapy and Oncology}, 141:192--199, 2019.

\bibitem{frangi1998multiscale}
Alejandro~F Frangi, Wiro~J Niessen, Koen~L Vincken, and Max~A Viergever.
\newblock Multiscale vessel enhancement filtering.
\newblock In {\em International conference on medical image computing and
  computer-assisted intervention}, pages 130--137. Springer, 1998.

\bibitem{frid2019endotracheal}
Maayan Frid-Adar, Rula Amer, and Hayit Greenspan.
\newblock Endotracheal tube detection and segmentation in chest radiographs
  using synthetic data.
\newblock In {\em International Conference on Medical Image Computing and
  Computer-Assisted Intervention}, pages 784--792. Springer, 2019.

\bibitem{he2016deep}
Kaiming He, Xiangyu Zhang, Shaoqing Ren, and Jian Sun.
\newblock Deep residual learning for image recognition.
\newblock In {\em Proceedings of the IEEE conference on computer vision and
  pattern recognition}, pages 770--778, 2016.

\bibitem{heimann2009statistical}
Tobias Heimann and Hans-Peter Meinzer.
\newblock Statistical shape models for 3d medical image segmentation: a review.
\newblock {\em Medical image analysis}, 13(4):543--563, 2009.

\bibitem{ibragimov2014shape}
Bulat Ibragimov, Bo{\v{s}}tjan Likar, Franjo Pernu{\v{s}}, and Toma{\v{z}}
  Vrtovec.
\newblock Shape representation for efficient landmark-based segmentation in
  3-d.
\newblock {\em IEEE Transactions on Medical Imaging}, 33(4):861--874, 2014.

\bibitem{kholiavchenko2020contour}
Maksym Kholiavchenko, Ilyas Sirazitdinov, K Kubrak, R Badrutdinova, Ramil
  Kuleev, Yixuan Yuan, Tomaz Vrtovec, and Bulat Ibragimov.
\newblock Contour-aware multi-label chest x-ray organ segmentation.
\newblock {\em International Journal of Computer Assisted Radiology and
  Surgery}, 15(3):425--436, 2020.

\bibitem{korkinof2018high}
Dimitrios Korkinof, Tobias Rijken, Michael O'Neill, Joseph Yearsley, Hugh
  Harvey, and Ben Glocker.
\newblock High-resolution mammogram synthesis using progressive generative
  adversarial networks.
\newblock {\em arXiv preprint arXiv:1807.03401}, 2018.

\bibitem{lorenz1997multi}
Cristian Lorenz, I-C Carlsen, Thorsten~M Buzug, Carola Fassnacht, and
  J{\"u}rgen Weese.
\newblock Multi-scale line segmentation with automatic estimation of width,
  contrast and tangential direction in 2d and 3d medical images.
\newblock In {\em CVRMed-MRCAS'97}, pages 233--242. Springer, 1997.

\bibitem{mao2017least}
Xudong Mao, Qing Li, Haoran Xie, Raymond~YK Lau, Zhen Wang, and Stephen
  Paul~Smolley.
\newblock Least squares generative adversarial networks.
\newblock In {\em Proceedings of the IEEE International Conference on Computer
  Vision}, pages 2794--2802, 2017.

\bibitem{miyato2018spectral}
Takeru Miyato, Toshiki Kataoka, Masanori Koyama, and Yuichi Yoshida.
\newblock Spectral normalization for generative adversarial networks.
\newblock {\em arXiv preprint arXiv:1802.05957}, 2018.

\bibitem{neff2018data}
Thomas Neff.
\newblock {\em Data augmentation in deep learning using generative adversarial
  networks}.
\newblock PhD thesis, Master’s thesis, Graz University of Technology, Graz,
  Austria, 2018.

\bibitem{nikolenko2019synthetic}
Sergey~I Nikolenko.
\newblock Synthetic data for deep learning.
\newblock {\em arXiv preprint arXiv:1909.11512}, 2019.

\bibitem{cvat}
Foundation OpenCV.
\newblock opencv/cvat.

\bibitem{patil2013medical}
Dinesh~D Patil and Sonal~G Deore.
\newblock Medical image segmentation: a review.
\newblock {\em International Journal of Computer Science and Mobile Computing},
  2(1):22--27, 2013.

\bibitem{prokopenko2019synthetic}
Denis Prokopenko, Jo{\"e}l~Valentin Stadelmann, Heinrich Schulz, Steffen
  Renisch, and Dmitry~V Dylov.
\newblock Synthetic ct generation from mri using improved dualgan.
\newblock {\em arXiv preprint arXiv:1909.08942}, 2019.

\bibitem{prokopenko2019unpaired}
Denis Prokopenko, Jo{\"e}l~Valentin Stadelmann, Heinrich Schulz, Steffen
  Renisch, and Dmitry~V Dylov.
\newblock Unpaired synthetic image generation in radiology using gans.
\newblock In {\em Workshop on Artificial Intelligence in Radiation Therapy},
  pages 94--101. Springer, 2019.

\bibitem{radford2015unsupervised}
Alec Radford, Luke Metz, and Soumith Chintala.
\newblock Unsupervised representation learning with deep convolutional
  generative adversarial networks.
\newblock {\em arXiv preprint arXiv:1511.06434}, 2015.

\bibitem{ronneberger2015u}
Olaf Ronneberger, Philipp Fischer, and Thomas Brox.
\newblock U-net: Convolutional networks for biomedical image segmentation.
\newblock In {\em International Conference on Medical image computing and
  computer-assisted intervention}, pages 234--241. Springer, 2015.

\bibitem{sato19973d}
Yoshinobu Sato, Shin Nakajima, Hideki Atsumi, Thomas Koller, Guido Gerig,
  Shigeyuki Yoshida, and Ron Kikinis.
\newblock 3d multi-scale line filter for segmentation and visualization of
  curvilinear structures in medical images.
\newblock In {\em CVRMed-MRCAS'97}, pages 213--222. Springer, 1997.

\bibitem{shit2020clDice}
Suprosanna Shit, Johannes~C Paetzold, Anjany Sekuboyina, Andrey Zhylka, Ivan
  Ezhov, Alexander Unger, Josien~PW Pluim, Giles Tetteh, and Bjoern~H Menze.
\newblock cldice--a topology-preserving loss function for tubular structure
  segmentation.
\newblock {\em arXiv preprint arXiv:2003.07311}, 2020.

\bibitem{subramanian2019automated}
Vaishnavi Subramanian, Hongzhi Wang, Joy~T Wu, Ken~CL Wong, Arjun Sharma, and
  Tanveer Syeda-Mahmood.
\newblock Automated detection and type classification of central venous
  catheters in chest x-rays.
\newblock In {\em International Conference on Medical Image Computing and
  Computer-Assisted Intervention}, pages 522--530. Springer, 2019.

\bibitem{wang2017chestx}
Xiaosong Wang, Yifan Peng, Le Lu, Zhiyong Lu, Mohammadhadi Bagheri, and
  Ronald~M Summers.
\newblock Chestx-ray8: Hospital-scale chest x-ray database and benchmarks on
  weakly-supervised classification and localization of common thorax diseases.
\newblock In {\em Proceedings of the IEEE conference on computer vision and
  pattern recognition}, pages 2097--2106, 2017.

\bibitem{wolterink2016evaluation}
Jelmer~M Wolterink, Tim Leiner, Bob~D De~Vos, Jean-Louis Coatrieux, B~Michael
  Kelm, Satoshi Kondo, Rodrigo~A Salgado, Rahil Shahzad, Huazhong Shu, Miranda
  Snoeren, et~al.
\newblock An evaluation of automatic coronary artery calcium scoring methods
  with cardiac ct using the orcascore framework.
\newblock {\em Medical physics}, 43(5):2361--2373, 2016.

\bibitem{yi2020automatic}
Xin Yi, Scott Adams, Paul Babyn, and Abdul Elnajmi.
\newblock Automatic catheter and tube detection in pediatric x-ray images using
  a scale-recurrent network and synthetic data.
\newblock {\em Journal of digital imaging}, 33(1):181--190, 2020.

\bibitem{yi2020computer}
Xin Yi, Scott~J Adams, Robert~DE Henderson, and Paul Babyn.
\newblock Computer-aided assessment of catheters and tubes on radiographs: How
  good is artificial intelligence for assessment?
\newblock {\em Radiology: Artificial Intelligence}, 2(1):e190082, 2020.

\bibitem{zhu2017unpaired}
Jun-Yan Zhu, Taesung Park, Phillip Isola, and Alexei~A Efros.
\newblock Unpaired image-to-image translation using cycle-consistent
  adversarial networks.
\newblock In {\em Proceedings of the IEEE international conference on computer
  vision}, pages 2223--2232, 2017.

\end{thebibliography}
}

\end{document}